\documentclass[12pt]{article}
\usepackage{graphicx}
\usepackage{amsmath}
\usepackage{amssymb}

\newcommand\pubnumber{DO-TH 14/27}
\newcommand\pubdate{\today}

\def\dortmund{
Institut f\"ur Physik,
Technische Universit\"at Dortmund,
Dortmund, Germany
}
\def\munich{
Excellence Cluster Universe,
Technische Universit\"at M\"unchen,
Garching, Germany
}
\def\supportMJ{\footnote{The work of M.J. is supported in part by the Bundesministerium f\"ur Bildung und Forschung (BMBF) and the DFG cluster of
excellence ``Origin and Structure of the Universe''.}}

\def\Title#1{\begin{center} {\Large #1 } \end{center}}
\def\Author#1{\begin{center}{ \sc #1} \end{center}}
\def\Address#1{\begin{center}{ \it #1} \end{center}}

\newcommand\pubblock{\rightline{\begin{tabular}{l} \pubnumber\\
         \pubdate  \end{tabular}}}
\newenvironment{Abstract}{\begin{quotation}  }{\end{quotation}}
\newenvironment{Presented}{\begin{quotation} \begin{center} 
             PRESENTED AT\end{center}\bigskip 
      \begin{center}\begin{large}}{\end{large}\end{center} \end{quotation}}
\def\Acknowledgements{\bigskip  \bigskip \begin{center} \begin{large}
             \bf ACKNOWLEDGEMENTS \end{large}\end{center}}

\def\beq{\begin{equation}}
\def\eeq#1{\label{#1}\end{equation}}
\def\eeqn{\end{equation}}

\def\beqa{\begin{eqnarray}}
\def\eeqa#1{\label{#1}\end{eqnarray}}
\def\eeqan{\end{eqnarray}}

\let\bar=\overbar

\def\ie{{\it i.e.}}

\def\Dslash{\not{\hbox{\kern-4pt $D$}}}
\def\dslash{\not{\hbox{\kern-2pt $\del$}}}

\def\msb{{\bar{\ssstyle M \kern -1pt S}}}

\begin{document}
\begin{titlepage}
\pubblock

\vfill
\Title{Summary of WG 4:\\ Mixing and mixing-related CP violation in the $B$ system}
\vfill
\Author{Yasmine Sara Amhis$^a$\index{Amhis, Y.}, Tagir Aushev$^b$\index{Aushev, T.}, Martin Jung$^{c,d}$\index{Jung, M.}\supportMJ}
\Address{${}^{a}$LAL, Universit\'e Paris-Sud, CNRS/IN2P3, Orsay France\\
${}^{b}$Moscow Institute of Physics and Technology, MIPT, Moscow Russia\\
${}^{c}$\munich\\ ${}^{d}$\dortmund
}
\vfill
\begin{Abstract}
We present the summary of the working group on $B$ mixing and the related CP violation at the CKM 2014 workshop. The contributions reflect the
experimental and theoretical progress in the field over the last two years since the last CKM workshop.
\end{Abstract}
\vfill
\begin{Presented}
8th International Workshop on the CKM Unitarity Triangle\\
8-12 September 2014, Vienna, Austria
\end{Presented}
\vfill
\end{titlepage}
\def\thefootnote{\fnsymbol{footnote}}
\setcounter{footnote}{0}

\section{Introduction}

Mixing and CP violation in the $B$ system have been essential in establishing the Cabibbo-Kobayashi-Maskawa (CKM) picture of flavour- and CP violation
in the Standard Model (SM) \cite{CKM}, as exemplified by the successful fits to the Unitarity Triangle
(UT)~\cite{Charles:2004jd,Ciuchini:2000de,Bevan:2014cya,TalkDescotesGenon,TalkBona},\footnote{The long-standing tension between the inclusive and
exclusive extraction of the CKM matrix elements $V_{ub}$ and $V_{cb}$ is discussed in more detail in WG~2~\cite{SummaryWG2}.} and continue to provide important constraints on SM
parameters and new physics (NP) models, some of which are discussed in the following.

The present article summarizes 21 contributions, grouped according to the section topics ``$B$ mixing and $b$-hadron lifetimes'',
``Determination of the mixing phases $\phi_{d,s}$'', and ``Determination of the UT angles $\alpha(\phi_2)$ and $\gamma(\phi_3)$''.

\section{$\mathbf B$ mixing and $\mathbf b$-hadron lifetimes}

The mixing of neutral mesons with their antiparticles can be characterized by three quantities: $|M_{12}|$, $|\Gamma_{12}|$, and
$\phi_{12}=\rm{arg}(-M_{12}/\Gamma_{12})$, all related to measurable quantities, see Ref.~\cite{Lenz:2014jya} for details and further
references.\footnote{Note that the convention used here is slightly different from \cite{Lenz:2014jya}: the additional index '12' is added here,
because $\phi_{d,s}$ is used below for the phase appearing in time-dependent $B_{d,s}$ decays.} $M_{12}$ is related to the dispersive part of the
transition amplitude $\langle B|\mathcal H|\bar B\rangle$; this quantity is sensitive to heavy particles in the loop, \ie{} the top quark in the SM
and potential new particles in NP models. On the other hand, $\Gamma_{12}$ is determined from the absorptive part of the same amplitude, and thereby
less sensitive to NP.

The calculation of these quantities in the SM is facilitated by large hierarchies: the fact that $M_{12}$ is dominated by contributions from heavy
internal particles allows for using an effective field theory with only one effective four-quark operator, the coefficient of which can be computed
reliably in perturbation theory \cite{Inami:1980fz,Buras:1990fn}. The corresponding hadronic matrix element is accessible on the lattice, as are the
ones for the operators appearing in NP models, with uncertainties at the few-pe:rcent level, see Ref.~\cite{Carrasco:2014nda} for a recent review. The
calculation of $\Gamma_{12}$ is more involved, since lighter degrees of freedom are relevant; a second operator product expansion is necessary, the so-called
heavy quark expansion (HQE), which exploits a second hierarchy, $m_b\gg \Lambda_{\rm QCD}$, see \emph{e.g.} Ref.~\cite{Lenz:2014jha} for details and
references. The precision of these calculations is presently around $20\%$, limited by the remaining non-perturbative parameters \cite{Lenz:2014jya}.
Several discrepancies between experimental results and HQE predictions used to question the validity of the HQE; these have been resolved over the
last years, most recently with high-precision results from the LHC experiments, see Refs.~\cite{Lenz:2014jya,TalkDordei} and the discussion
below, yielding a very consistent picture. A remaining puzzle is the measurement of a relatively large like-sign dimuon asymmetry (LSDA) by the D0
collaboration~\cite{LSDA,Abazov:2013uma}, discussed below and in \cite{TalkNierste,Nebot:2014yqa,TalkBertram}.
 
From an experimental point of view, the determination of $b$-hadron lifetimes has evolved from early measurements by the CLEO and LEP experiments
to precision measurements by the $B$-factories and the Tevatron experiments;  
recently, the LHCb Collaboration further improved the precision of these earlier measurements significantly for all weakly decaying $b$-hadrons, up
to an order of magnitude in some cases~\cite{TalkDordei}. This was possible using about 3 fb$^{-1}$ of collected proton-proton collision data
delivered by the LHC at 7 and 8~TeV center-of-mass energy in 2011 and 2012, respectively. 
Hadron colliders benefit from two major advantages, the large production cross-section of $b$-quarks and the production of all species of $b$-hadrons
in the hadronization process.  However, the challenge for them is to cope with an extremely large data rate. 
The trigger system reduces this rate to amounts that can be written on disk.
Throughout the selection of the $b$-candidates of interest in the trigger system and during further processing, for example the track reconstruction,
some quantities like the impact parameter of the $b$-candidate will distort and/or bias the distribution of the decay-time acceptance.
To overcome these experimental issues, two possible approaches can be adopted. The first one is to perform absolute
lifetime measurements, which is harder experimentally. 
While typically modes with large branching fractions can be used, leading to a good statistical precision, this approach requires an excellent
knowledge of all the  small systematic effects that contribute to the distortion of the observable of interest, \emph{i.e.} the lifetime.
The alternative is to perform relative lifetime measurements. They rely on the fact that most of the systematic effects will cancel in the lifetime
ratio. LHCb has applied both approaches and the various analyses are summarized in the following.

With 1~fb$^{-1}$ of collected data LHCb performed the most precise absolute lifetime measurements to date for $B^0_{d,s}$, $B^-$ and $\Lambda_b^0$,
exploring decays involving a $J/\Psi$~\cite{LHCb:1402.2554}. The modes in question are: $B^+ \to J/\psi K^+, B^0 \to J/\psi K^*(892)^0,  B^0 \to
J/\psi K^0_S, B^0_s \to J/\psi \phi$ and $\Lambda^0_b \to J/\psi \Lambda$. The results are reported in Table~\ref{tab:sfit_fit_results}.
 Additionally, as suggested in Ref.~\cite{TGershon2010}, a combined analysis of both $B^0 \to J/\psi K^*(892)^0$ and $B^0 \to J/\psi K^0_S$ together
 with the knowledge of
the mixing phase $\beta= (21.5^{+0.8}_{-0.7})^\circ$~\cite{Agashe:2014kda} allows for the measurement of $\Delta \Gamma_d/\Gamma_d$, which can be
 used as a probe for NP searches as \emph{e.g.} argued in Ref.~\cite{Lenz:2014jya}. However, the measured value was found to be $\Delta \Gamma_d/\Gamma_d
= -0.044 \pm 0.025 \rm{\scriptstyle(stat)} \pm0.011 \rm{\scriptstyle(syst)}$, consistent with SM predictions.
\begin{table}[htb]
\centerline{
\begin{tabular}{lc}
        \hline \hline
	Lifetime		&	Value [ps]\\
	\hline
	$\tau_{B^+ \to J/\psi K^+}$	  &	1.637 $\pm$ 0.004 $\pm$ 0.003 \\
	$\tau_{B^0 \to J/\psi K^*(892)^0}$&	1.524 $\pm$ 0.006 $\pm$ 0.004 \\
	$\tau_{B^0 \to J/\psi K^0_S}	$&	1.499 $\pm$ 0.013 $\pm$ 0.005 \\
	$\tau_{\Lambda^0_b \to J/\psi \Lambda}$	  &	1.415 $\pm$ 0.027 $\pm$ 0.006 \\
	$\tau_{B^0_s \to J/\psi \phi}	$&	1.480 $\pm$ 0.011 $\pm$ 0.005 \\
	\hline
\end{tabular}
}
\caption{\small Fit results for the $B^+$, $B^0$, $B^0_s$ mesons and $\Lambda^0_b$ baryon lifetimes measured at LHCb~\cite{LHCb:1402.2554}. The first
uncertainty is statistical and the second is systematic.}
\label{tab:sfit_fit_results}
\end{table}

The abundant, yet previously unobserved $\Lambda^0_b \to J/\psi p K $ mode provides the most precise measurement of the lifetime ratio
$\tau_{\Lambda_b}/\tau_{B^0}$ to date. Using 3~fb$^{-1}$, the lifetime of $\Lambda_b^0$  was measured to be $1.479\pm 0.009  \rm{\scriptstyle(stat)}
\pm 0.010  \rm{\scriptstyle(syst)}$~ps~\cite{LHCb:1307.2476}. This measurement is compatible with other recent measurements of this quantity (like the
one above, see Ref.~\cite{Agashe:2014kda} and references therein) as well as with the HQE prediction~\cite{Lenz:2014jha}. Therefore the long-standing puzzle of this
ratio being measured lower than expected from theory has been resolved.

In two LHCb analyses with 3~fb$^{-1}$ the relative lifetimes of $b$-baryons containing strange quarks were improved and confronted with their
theoretical predictions from HQE.
The $\Xi_b^{0}$ lifetime was measured for the first time~\cite{LHCb:arXiv:1405.7223}, using the hadronic decay mode $\Xi_b^{0}\rightarrow
\Xi_c^{+}\pi^{-}$, resulting in $\tau_{\Xi_b^0\rightarrow \Xi_c^{+}\pi^{-}}=1.477 \pm 0.026\rm{\scriptstyle(stat)} \pm 0.014 \rm{\scriptstyle(syst)}
\pm 0.013 \rm{\scriptstyle(input)} $~ps. The $\Xi_b^{-}$ and $\Omega_b^{-}$ lifetimes were measured reconstructing a dimuon ($J/\psi$) and a hyperon
($\Xi^{-},\Omega^{-}$) in the final state~\cite{LHCb:arXiv:1405.1543} and were found to be $\tau_{\Xi^-_b\to
J/\Psi\Xi^-}=1.55^{+0.10}_{-0.09}\rm{\scriptstyle(stat)} \pm 0.03 \rm{\scriptstyle(syst)}$~ps and $\tau_{\Omega_b^-\to
J/\Psi\Omega^{-}}=1.54^{+0.26}_{-0.21}\rm{\scriptstyle(stat)} \pm0.05 \rm{\scriptstyle(syst)}$~ps. The results agree again very well with
the HQE estimate $\tau_{\Xi_b^0}/\tau_{\Xi_b^+}=0.95\pm0.06$~\cite{Lenz:2014jha} and are in the case of the $\Omega_b$ within the range of earlier
estimates. %

\begin{figure}[htb]
\centering
\includegraphics[height=7.3cm]{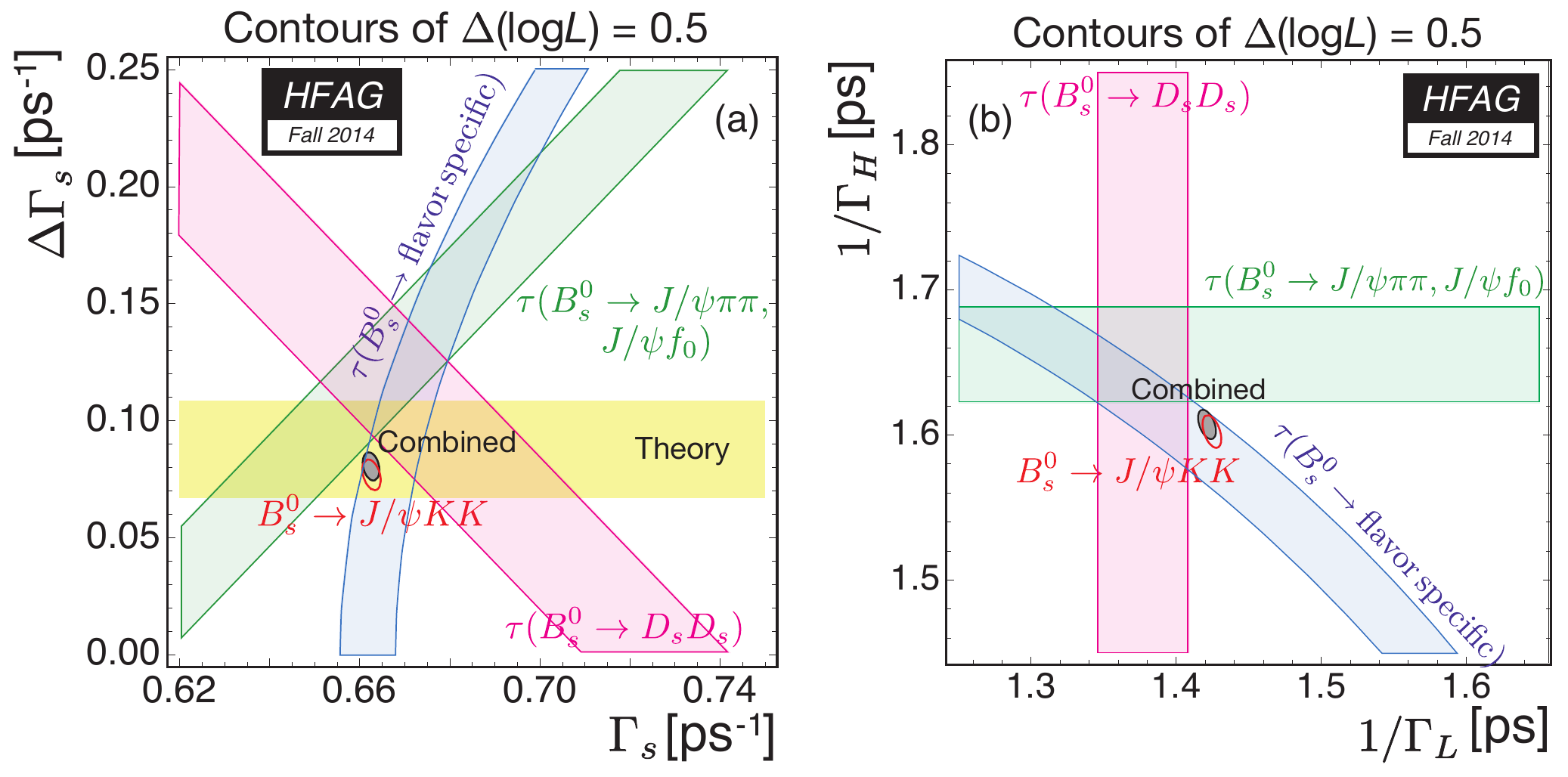}
\caption{Available measurements of $\Gamma_s,\Delta\Gamma_s$ in the $\Delta\Gamma_s$ vs. $\Gamma_s$ (left) and $\Gamma_L$ vs.
$\Gamma_H$ (right) planes.}
\label{fig:DG}
\end{figure}

While in the $B^0$ system the width difference between the heavy and light mass eigenstates is predicted to be tiny and
measured compatible with zero, see above, this is not the case for the $B^0_s$ system. 
This affects branching ratio measurements and opens the possibility to access the rate asymmetry and the width difference via the effective
lifetime~\cite{Fleischer:2011cw}, 
\begin{equation}
\tau_f^{\rm eff} = \frac{\int_0^\infty\!\! dt\,\,t\,\langle\Gamma(B_s(t)\to f)\rangle}{\int_0^\infty\!\! dt\,\langle\Gamma(B_s(t)\to
f)\rangle}=\frac{\tau_{B_s}}{1-y_s^2}\frac{1+2\mathcal A_{\Delta\Gamma}^f y_s+y_s^2}{1+\mathcal A_{\Delta\Gamma}^f y_s} \,,
\end{equation}
where $y_s=\Delta\Gamma_s/(2\Gamma_s)$.
In the SM decays such as $B_s^0\rightarrow K^+K^-$ and $B_s^0\rightarrow D_s^+D_s^-$ have tiny CP asymmetries, which is reflected in the
predictions  ${\cal A}_{\Delta\Gamma}^{B_s\to K^+K^-}=-0.97^{+0.004}_{-0.009}$ \cite{arXiv:1011.1096} and ${\cal A}_{\Delta\Gamma}^{B_s\to
D_s^+D_s^-}=-1+\mathcal O(10^{-3})$~\cite{Jung:2014jfa}. Therefore the measurement of the effective lifetime of these modes is equivalent to measuring
$\Gamma_L$. The lifetimes of both channels were measured using 1 and 3~fb$^{-1}$, respectively, and found to be $\tau_{B_s\to K^+K^-}^{\rm
eff}=1.407 \pm 0.016\rm{\scriptstyle(stat)} \pm 0.007 \rm{\scriptstyle(syst)}$~ps and $\tau_{B_s\to D_s^+D_s^-}^{\rm eff}=1.379 \pm
0.026\rm{\scriptstyle(stat)} \pm 0.017 \rm{\scriptstyle(syst)}$~ps. Finally, in flavour-specific  $B_s^0$  decays $\mathcal A_{\Delta \Gamma}^{fs}=0$
holds, yielding a direct measurement of $\Gamma_s^{-1}$ to first order in $y_s$. 
In particular, the measurements of the abundant hadronic $B_s^0\rightarrow D_s\pi$ and semi-leptonic $B_s^0\to D_s \mu \nu$ modes have been updated
by both the LHCb and the D0 experiments~\cite{LHCb:arXiv:1407.5873,D0:arXiv:1410.1568}, reaching a statistical uncertainty of $\mathcal O(10)$~fs. 
The global picture of the effective lifetime measurements in the $B_s^0$ system, depicted in Fig.~\ref{fig:DG}, shows good consistency.
The measurement of $\Gamma_s,\Delta \Gamma_s$ from $B_s\to J/\Psi KK$ decays, dominating the average, will be discussed later in the context of
the determination of the mixing angle $\phi_s$. 

The mixing structure described in the beginning is very general, it holds also in the presence of NP. A simple model-independent parametrization of NP
effects then reads $M_{12}^q = M_{12}^{q, \rm SM}\Delta_q$, with complex parameters $\Delta_q$, while $\Gamma_{12}$ used to be considered SM-like
\cite{Lenz:2006hd}. Detailed analyses of this type have been carried out over the years \cite{NPBmixing}; at the moment, the results are perfectly
compatible with the SM~\cite{Lenz:2014jya,Bevan:2014cya,TalkDescotesGenon}. Albeit room is left for NP influence of about $20\%$, this translates into
stringent limits on the generic scales of NP operators, typically much higher than directly accessible at colliders
\cite{Bevan:2014cya,TalkDescotesGenon}.
Given the absence of large NP effects in $M_{12}$, the increasing experimental precision for the observables involved, and the fact that the LSDA
cannot be explained in a fit with $\Delta_q$, only, there has been an increased interest in considering NP in $\Gamma_{12}^q$. Regarding $\Delta
\Gamma_q$, for $q=s$ the relative influence is already bound to be below approximately $30\%$ \cite{Bobeth:2011st}, while it can still be large for
$q=d$ \cite{Lenz:2014jya,Bobeth:2014rda}; $b\to d\tau^+\tau^-$ transitions are especially interesting in that respect and could
still show a large enhancement.

The CP asymmetry in mixing of neutral B decays can be measured via decays to flavour-specific final states (typically semileptonic ones), 
\begin{equation}
\label{eq:asl}
a_{\rm fs} = a_{\rm sl} = \frac{\Gamma[\overline{B} (\rightarrow B) \rightarrow f] - \Gamma[B (\rightarrow \overline{B}) \rightarrow
\overline{f}]}{\Gamma[\overline{B} (\rightarrow B) \rightarrow f] + \Gamma[B (\rightarrow \overline{B}) \rightarrow
\overline{f}]}=\left|\frac{\Gamma_{12}^q}{M_{12}^q}\right|\sin\phi_{12}^q\,.
\end{equation}
It is expected to be very small in the SM, below the present level of experimental precision. It could however be enhanced in the presence of NP
entering the mixing amplitude.

\begin{figure}[htb]
\centering
\includegraphics[height=8cm]{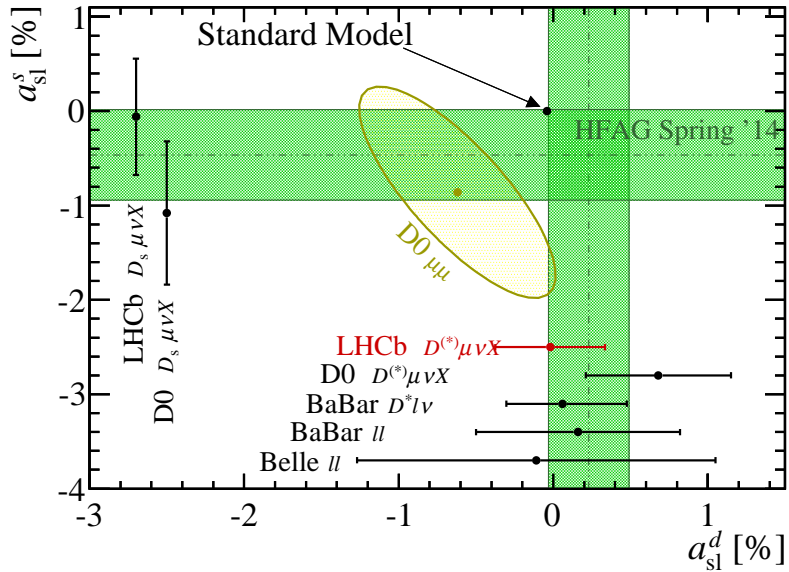}
\caption{Overview of the measurements of the CP asymmetry in mixing of $B_{d,s}$ mesons.
The preliminary measurement from the BaBar collaboration, $ a_{\rm sl}^{d} = (-3.9\pm 3.5({\rm stat}) \pm 1.9({\rm syst})) \cdot
10^{-3}$~\cite{TalkCheng,Lees:2014kep}, presented at this conference, is included and considered in the average. The bands correspond to the
average of the pure $a_{sl}^d$ and $a_{sl}^s$ measurements, which are in conflict with the D0 dimuon result.}
\label{fig:asl}
\end{figure}

Recently, the LHCb experiment measured  the asymmetries in both $B^{0}$ and $B_s^0$ systems~\cite{LGrillo:2014,LHCb:arXiv:1409.8586}:
$a^{s}_{sl} = [-0.06  \pm0.50   \rm{\scriptstyle(stat)}  \pm  0.63\rm{\scriptstyle(syst)} ]\%$ and
$a^{d}_{sl} = [-0.02  \pm0.19   \rm{\scriptstyle(stat)}  \pm  0.30\rm{\scriptstyle(syst)} ]\%$.
Also the BaBar experiment presented at this workshop a new measurement 
using dimuon events from the full BaBar dataset of $471\times10^6$ $B\bar B$ pairs, yielding 
$a^{d}_{sl} = (-3.9\pm3.5({\rm stat})\pm1.9({\rm syst}))\times10^{-3}$~\cite{TalkCheng,Lees:2014kep}. Both LHCb and BaBar measurements show an
excellent agreement with the SM predictions. However, as shown in Fig.~\ref{fig:asl}, some tension still remains with the measurement of the LSDA
performed by D0~\cite{Abazov:2013uma}, $A_{CP}=(-0.235\pm0.064 \rm{\scriptstyle(stat)}\pm0.055\rm{\scriptstyle(stat)})\%$, 3.6 standard deviations
away from the SM.
Regarding this measurement, in Ref.~\cite{Borissov:2013wwa} the interesting observation has been made that not only the flavour-specific CP
asymmetries contribute, but also mixing-induced CP asymmetries, contributing proportional to $\Delta \Gamma_q$. A more detailed
analysis~\cite{TalkNierste} shows that the effect is about $50\%$ smaller than estimated in Ref.~\cite{Borissov:2013wwa}, yielding a \emph{larger}
tension than quoted above. While this effect does have an influence on the LSDA, it seems too small to explain the measurement within the SM, thereby
still hinting at a NP explanation. However, even in NP models it is rather difficult to achieve a large enhancement without violating other constraints.
Interestingly, many constraints can be avoided using the fact that the dependence on the individual contributions to $\Gamma_{12}$ is different for
the LSDA, the flavour-specific CP asymmetry, and $\Delta \Gamma$ \cite{TalkNierste}. One class of models where an enhancement is related to a
potential non-unitarity of the CKM matrix has been discussed in \cite{Nebot:2014yqa,Botella:2014qya} and references therein. However, also here the
enhancement is limited, especially by the constraints from $V_{ub}$ and $S_{\rm CP}(B_s\to J/\psi \phi)$, yielding values of at most $10^{-3}$ for the
LSDA.

\section{Determination of the mixing phases $\boldsymbol{\phi_{d,s}}$\label{sec::phids}}

The precision extractions of the mixing phases $\phi_{d,s}$ aim at the precise knowledge of these SM parameters, but more importantly the discovery of
potential NP contributions. These phases can be cleanly determined in tagged time-dependent analyses of $b\to c\bar c s$
transitions~\cite{Bigi:1981qs}. The advantage of this class of decays is that the amplitude is dominated to good approximation by its contribution
proportional to $\lambda_{cs}\equiv V_{cb}V_{cs}^*$, while the subleading parts, usually jointly dubbed \emph{penguin pollution}, are not only
CKM-suppressed by $\lambda_{us}/\lambda_{cs}\approx 2\%$, but are also expected to have smaller hadronic matrix elements; however, this latter
suppression is hard to quantify for the decays in question. 
With the time-dependent CP asymmetry of a decay $\mathcal D$ into a CP eigenstate given as 
\begin{eqnarray}\label{eq::DefCPA}
a_{\rm CP}(\mathcal D;t)\! \equiv\! \frac{\Gamma(\mathcal D;t)-\Gamma(\bar{\mathcal D};t)}{\Gamma(\mathcal D;t)+\Gamma(\bar{\mathcal D};t)}
=\! \frac{S_{\rm CP}(\mathcal D)\sin(\Delta m t)-C_{\rm CP}(\mathcal D)\cos(\Delta m t)}{\cosh(\Delta \Gamma t/2)+A_{\Delta \Gamma}(\mathcal
D)\sinh(\Delta \Gamma t/2)}\,,
\end{eqnarray}
the corrections of the resulting (schematic) relations
\begin{equation}\label{eq:goldenmodes}
S_{\rm CP}(B_{d,s}\to J/\psi X)\simeq\pm\sin\phi_{d,s}\,,\qquad C_{\rm CP}(B_{d,s}\to J/\psi X)\simeq0\,, 
\end{equation}
are estimated to be of the order $\mathcal{O}(10^{-3})$, only \cite{PenEstimates}; it is, however, notoriously difficult to actually calculate the
relevant matrix elements, and non-perturbative enhancements cannot be excluded. Given the absence of large NP effects, as inferred already \emph{e.g.}
from Refs.~\cite{Charles:2004jd,Ciuchini:2000de,NPBmixing}, but for the decays in question also from the compatibility of the recent
measurements~\cite{Aaij:2014zsa,Aad:2014cqa,Khachatryan:2015nza} with the SM, only the quantitative control of these subleading contributions will
allow to fully exploit the precision measurements from the LHC experiments and Belle~II.

To gain control over these contributions, typically flavour-symmetry relations are used, where the
unknown matrix elements can be extracted from data. When using the $U$-spin subgroup of $SU(3)$, relating down and strange quarks, usually one
``control mode'' is used, where the relative influence of the penguin matrix elements is larger
\cite{Fleischer:1999nz,PenUspin,Faller:2008zc,Faller:2008gt}. Experimentally, the disadvantage of these modes is that their rates are suppressed by
$\lambda^2\sim5\%$. Theoretically, there are mainly two difficulties: firstly, since only one additional mode is used, it is not possible to control
penguin pollution and $U$-spin breaking contributions simultaneously. The necessary assumption for the latter can lead to a bias in the extraction of
the penguin shift \cite{Feldmann:2008fb,Jung:2012mp}. Secondly, for $B_s\to J/\psi\phi$, the control modes involve matrix elements of octet final
states, while the $\phi$ is a superposition of octet and singlet. The first issue has been addressed for the extraction of $\phi_d$ in
Ref.~\cite{Jung:2012mp} by extending the flavour symmetry to the full $SU(3)$ group, thereby including a full set of control modes. The additional
data allow to control the penguin shift and $SU(3)$-breaking contributions at the same time model-independently. The second issue has been estimated
to be a small effect~\cite{Gronau:2008hb}, but an extraction from data should be aimed for, which might be possible using the corresponding final
states with the $\omega$ meson.
The control of both effects seems therefore feasible in the future, allowing for precision extractions of $\phi_{d,s}$ even beyond the present level.

A new estimate for the penguin pollution in $B_d\to J/\psi K$ and $B_s\to J/\psi\phi$ has been
presented at this workshop~\cite{TalkFrings,Frings:2015eva}. Here it has been shown that the up-quark penguin contribution can be described in an
effective theory, resulting from an additional OPE in $1/q^2$ with $q^2\sim M_{J/\psi}^2$. This approach is again limited by the insufficient
knowledge of the corresponding hadronic matrix elements; estimating them in the $1/N_C$ approach, where $N_C=3$ is the number of colours, an upper
limit of $\Delta \phi_{d,s}\lesssim 1^\circ$ is obtained, consistent with the limits obtained from flavour symmetries described above.

Apart from the ``golden'' modes, $B_d\to J/\psi K_S$ and $B_s\to J/\psi\phi$, also the $B_s\to J/\psi f_0(980)$ decay was proposed to extract
$\phi_s$~\cite {BtoJPsif0prop}, $f_0(980)$ being the largest resonance in $B_s\to J/\psi\pi^+\pi^-$; since it is a scalar meson, this mode has the
advantage that no angular analysis is necessary, yielding a sensitivity similar to $B_s\to J/\psi\phi$.  Of concern in this case is the hadronic
nature of the $f_0(980)$, which has untypical characteristics for a simple $q\bar q$ meson (see \emph{e.g.} Ref.~\cite{ReviewsScalars} for recent
reviews), and its mixing with the $\sigma(f_0(500))$ resonance. Importantly, the hadronic features influence the decay dynamics, and specifically the
penguin contributions~\cite{Fleischer:2011au}. This renders a control-mode analysis of the type described above more complicated. In
Ref.~\cite{Aaij:2014siy} a pure tetraquark interpretation was excluded under the assumption of a vanishing mixing angle between $f_0(980)$ and
$\sigma$. Dropping this assumption and re-examining earlier constraints could re-open this possibility~\cite{Knegjens:2014fqa}. In any case the
hadronic nature of the $f_0(980)$ remains an open issue and the control of subleading contributions to a comparable level as in $B_s\to J/\psi\phi$
seems hard to achieve.

Experimentally, the decay mode $B_s^0\to J/\psi\phi$  is not only used to measure $\phi_s$, but also $\Delta \Gamma_s$ and $\Gamma_s$.  A
time-dependent analysis is necessary to separate the CP-odd and CP-even components in this decay, as also discussed at this
conference~\cite{GBorissov:2014,JPazzini:2014,WKanso:2014}.
The experimental techniques are very similar across the three experiments ATLAS, CMS, and LHCb, using an unbinned maximum likelihood. The fits are
multi-dimensional, including for example the invariant mass of the $J/\psi\phi$ system, the decay time and the 3 angles in the helicity basis (LHCb)
or transversity basis (CMS, ATLAS). In all cases, flavour algorithms based on different taggers like the same-side Kaon tagger at LHCb and the
opposite-side lepton tagger at CMS and ATLAS are used to identify the flavour of the $B_s^0$ meson when it was produced in the LHC collisions.

\begin{table}
\centerline{
\begin{tabular}{l l l l}\hline\hline
Exp.        & $\phi_s/{\rm rad}$        & $\Delta \Gamma_s/{\rm ps}^{-1}$   
& Comments\\\hline
ATLAS       & \phantom{-}0.12(25)(11)   & 0.053(21)(9)                                             & $B_s^0\to J/\Psi KK$, $|\lambda|\equiv 1$
$4.9~{\rm fb}^{-1}$, \cite{GBorissov:2014,Aad:2014cqa}\\
CMS         & -0.08(10)(3)              & 0.095(13)(7) & $B_s^0\to J/\Psi KK$, $|\lambda|\equiv 1$, $20.0~{\rm fb}^{-1}$
\cite{JPazzini:2014,Khachatryan:2015nza}\\
LHCb        & \phantom{-}0.07(9)(1)     & $0.100(16)(3)$                           
    & $B_s^0\to J/\Psi KK$, $|\lambda|=0.94(3)(2)$, $1~{\rm fb}^{-1}$
\cite{WKanso:2014,Aaij:2013oba}\\
LHCb        & \phantom{-}0.07(7)(1)     & ---     & $B_s^0\to J/\Psi \pi\pi$, $|\lambda|=0.89(5)(1)$, $3~{\rm fb}^{-1}$
\cite{WKanso:2014,Aaij:2014dka}\\
\hline
\end{tabular}
}
\caption{\small \label{tab:resphis} Recent results for the mixing phase $\phi_s$  and width difference $\Delta\Gamma_s$.
The two uncertainties given are first the statistical and then the systematic one.}
\end{table}
The results from these analyses are collected in Table~\ref{tab:resphis}.\footnote{A couple of months after the CKM
workshop, LHCb updated their $B_s\to J/\Psi KK$ measurement using the full dataset collected during run~1. The results can be found in
Ref.~\cite{Aaij:2014zsa}} 
The $B_s\to J/\Psi KK$ analyses are focussed on the $B_s\to J/\Psi(\to \mu\mu)\phi(\to KK)$ chain. Regarding $B_s\to J/\Psi\pi\pi$, a full amplitude
analysis was done by the LHCb experiment to establish that the CP content of this channel is mainly CP-odd~\cite{Aaij:2014dka}. This fraction
was found to be higher than  $97.7\%$ at $95\%$~confidence level (CL).  Nevertheless, an angular analysis was used to include even this small
CP-even fraction precisely.

\begin{figure}[htb]
\centering
\includegraphics[height=8cm]{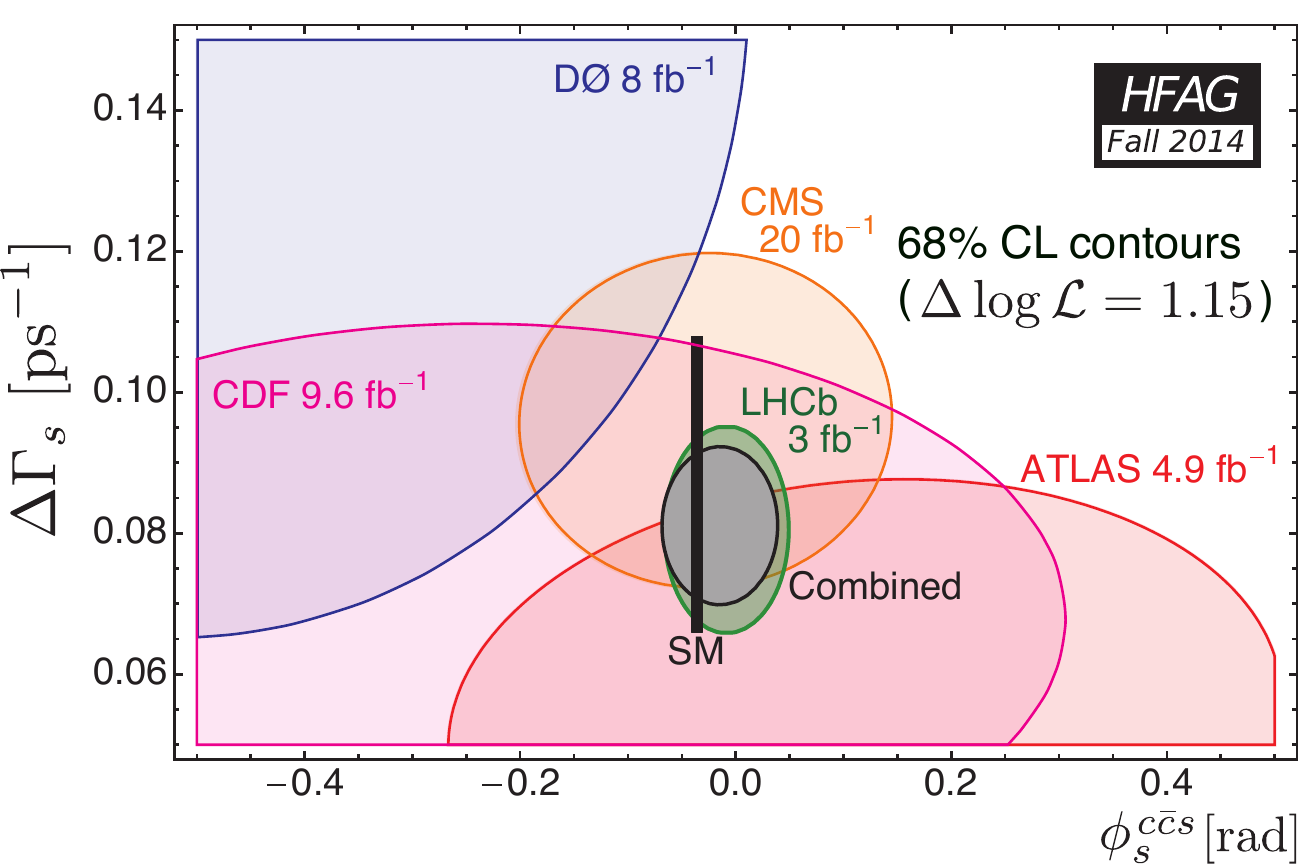}
\caption{Combination of the results for $\phi_s^{c\bar{c}s}$ vs. $\Delta\Gamma_s$ from various experiments. }
\label{fig:phisvsDG}
\end{figure}

The individual results and their average are displayed in Fig.~\ref{fig:phisvsDG}, showing good consistency. While the statistical uncertainty of
$\phi_s$ is still much larger than the one of the SM prediction, penguin pollution could already play a significant role. In that respect there has
been experimental progress as well, with measurements of the branching ratios of $B_s^0\to J/\Psi K_S^0$ and $B_s^0\to J/\Psi K^*$, together with
polarization fractions and CP-violating parameters for the latter by the LHCb collaboration~\cite{WKanso:2014,Aaij:2013eia,Aaij:2012nh,Aaij:2015mea}.
These modes can be used as control modes for $B^0\to J/\Psi K_S^0$~\cite{Faller:2008zc} and $B_s^0\to J/\Psi^0\phi$~\cite{Faller:2008gt}, respectively, using $SU(3)$ and, in the case of
$B_s\to J/\Psi K^*$, additional dynamical assumptions.

Another theoretically very clean mode sensitive to $\phi_s$ is $B_s^0\to D_s^+D_s^-$, providing an independent means to access this phase. Also here no
angular analysis is necessary, but experimentally the charmonium modes are easier to access. The penguin pollution in this mode is again very
difficult to calculate theoretically, but can be controlled with the means described above for the golden modes: the early proposal to use
$U$-spin~\cite{Fleischer:1999nz} has been extended to a full $SU(3)$ analysis~\cite{Jung:2014jfa}, presented at this conference~\cite{TalkSchacht},
allowing to control symmetry-breaking contributions model-independently. Additionally, $B^0\to D^{(*)}D^{(*)}$ decays allow for various other NP tests,
for example with quasi-isospin sumrules for branching ratios, and provide insights into QCD dynamics, like for instance weak
annihilation~\cite{TalkSchacht,Jung:2014jfa}.

The LHCb collaboration has measured the time-dependent CP asymmetry in $B_s^0\to D_s^+D_s^-$ for the first time, using $\Delta \Gamma_s$, $\Gamma_s$
and $\Delta m_s$ as external constraints~\cite{CFitzpatrick:2014,Aaij:2014ywt}. This constitutes the first measurement of  $\phi_s$ using a purely
hadronic final state. Using the full run~1 dataset, they report:
\begin{equation}
\phi_s =\; 0.02  \pm  0.17 \rm{\scriptstyle(stat)} \pm  0.02  \rm{\scriptstyle(syst)}\qquad\mbox{and}\qquad |\lambda| \;=\;  0.91^{+0.18}_{-0.25}
\rm{\scriptstyle(stat)}  \pm  0.02  \rm{\scriptstyle(syst)}\,,
\end{equation}
where the correlation between $\phi_s$ and $\lambda$ is 3 \%. Within the still sizable uncertainties this result is consistent with SM expectations.

The most recent analysis from the BaBar experiment of the decay mode $B^0\to D^{*+}D^{*-}$~\cite{2dstbabar} was also presented at this
conference~\cite{TMiyashita}.
This study uses the partial reconstruction technique,  where one of the final state mesons is fully reconstructed, while only the slow pion from the
decay of the second $D^{*\pm}$ meson is used. Neglecting the penguin contribution, which is a good approximation at this level of precision, the
following CP-violating parameters are obtained:
\begin{center}
$
  \begin{array}{lll}
    C_{\rm CP,+} &\;=\; & +0.15 \pm 0.09 \rm{(stat)} \pm 0.04 \rm{(syst)}, \\
    S_{\rm CP,+} &\;=\; & -0.49 \pm 0.18 \rm{(stat)} \pm 0.07 \rm{(syst)} \pm 0.04.\\
  \end{array}
$
\end{center}
The third uncertainty for the $S_{\rm CP,+}$ term is due to the input value of the CP-odd fraction, fixed to $R_\perp=0.158\pm0.029$, which is
necessary to extract from the effectively measured parameters $S_{\rm CP}$ and $C_{\rm CP}$, which involve admixtures of CP eigenstates, the ones for
the CP eigenstates which obey Eq.~\eqref{eq:goldenmodes} when neglecting penguin pollution. These results are consistent with SM expectations as well
as previous measurements.

Other processes sensitive to $\phi_{d,s}$ are $b\to s\bar qq$ transitions of $B_{d,s}$ mesons, where $q$ is a light quark. These modes are dominated
by penguin contributions in the SM, leading again to Eq.~\eqref{eq:goldenmodes}. Corrections to this relation are largely calculable in QCD
factorization and have been estimated in Ref.~\cite{Beneke:2005pu}. The induced shifts are larger than for $b\to c\bar cs$ transitions, since they are
determined by the ratio of the ``pollution'' and the leading amplitude which is suppressed. For the same reason the sensitivity to NP contributions is
generically larger for these modes, which constitutes the main interest in them.

At this workshop several analyses for this class of modes have been presented by the Belle
collaboration~\cite{TalkRitter,omegaKbelle,etaprimeKbelle}, namely time-dependent measurements for $B^0\to\omega K_S$,
$B^0\to\eta^\prime K^0$ and $B^0\to\eta K_S\gamma$.

In the decay mode $B^0\to\omega K_S^0$  the increased statistics of the full Belle dataset and improved reconstruction efficiency allowed to obtain
evidence for CP violation for the first time in this mode: the result reads~\cite{omegaKbelle} 
$A_{\rm CP}\equiv-C_{\rm CP}=-0.36\pm0.19\pm0.05$ and 
$S_{\rm CP}=+0.91\pm0.32\pm0.05$, which is $3.1\,\sigma$ away from 
the zero-CP-violation point and consistent with the SM expectation.

The decay mode $B^0\to\eta^\prime K^0$ includes two CP-eigenstates, the CP-even $B^0\to\eta^\prime K_S$ and the CP-odd $B^0\to\eta^\prime K_L$.
In total, $3541\pm91$ signal events are reconstructed in both decay modes.
The obtained results are consistent and are combined to yield the most precise measurement in this decay mode so far, $A_{{\rm
CP},f}=+0.03\pm0.05\pm0.04$ and $-\xi_f S_{{\rm CP},f}=+0.68\pm0.07\pm0.03$, where $\xi_f$ is the CP-eigenvalue of the corresponding final
state~\cite{etaprimeKbelle}. Again, no significant deviation from the SM expectation is observed.

Finally, an attempt is made to find CP violation in the decay mode $B^0\to\eta K_S\gamma$, however, no significant signal is observed.

\section{Determination of the UT angles $\boldsymbol{\alpha(\phi_2)}$ and $\boldsymbol{\gamma(\phi_3)}$}

Similarly to the UT angle $\beta(\phi_1)$ the angle $\alpha$ can be extracted from interference in $B^0(\bar B^0)$ decays, but using $b\to
d\bar uu$ transitions, specifically $B\to \pi\pi,\pi\rho,\rho\rho$ decays. If these were pure tree decays, they would be directly sensitive to
the combination $\phi_d+2\gamma\stackrel{\rm SM}{=}2\pi-2\alpha$. However, due to the different hierarchy of the CKM matrix
elements in $b\to d$ transitions the ``penguin pollution'' of these modes is much more significant than for $b\to s\bar cc$ transitions, and has to
be taken into account from the beginning. To disentangle the penguin and tree contributions, usually an isospin analysis is carried out in
$B\to\pi\pi$ and $B\to\rho\rho$~\cite{gronau}, relating the modes $\bar B^0\to h^+h^-$, $\bar B^0\to h^0h^0$ and $B^-\to h^-h^0$ via
$A^{+-}/\sqrt{2}+A^{00}=A^{+0},$ which can be presented as a triangle in the complex plane. The full information about the relative
size and (relative) phase of each amplitude can be extracted by measuring the corresponding rates and CP asymmetries.

The Belle experiment presented updated analyses for several of the relevant modes at this conference, namely 
 $B^0\to\pi^+\pi^-$, $B^0\to\pi^0\pi^0$ and $B^0\to\rho^0\rho^0$~\cite{Vanhoefer:2014mfa}. The first one has
a large branching fraction and high reconstruction efficiency, which allowed to reconstruct $2964\pm88$ signal events and to perform a time-dependent
measurement, yielding the most precise values of the CP-violation parameters to date: 
\begin{equation}
S_{\rm CP} = -0.64 \pm 0.08 \rm{(stat)} \pm 0.03 \rm{(syst)}\quad\mbox{and}\quad 
C_{\rm CP} = -0.33 \pm 0.06 \rm{(stat)} \pm 0.03 \rm{(syst)}.\\
\end{equation}
The branching ratio for $B\to\pi^0\pi^0$ is discussed in more detail in the summary of WG~III~\cite{Libby:2014cxa}; it is worth pointing out that the
new result, $BR(B^0\to\pi^0\pi^0)=(0.90\pm0.12\pm0.10)\times10^{-6}$, is significantly smaller than the previous Belle result~\cite{Abe:2004mp} and
shows a $\sim3\sigma$ tension with the BaBar one~\cite{Lees:2012mma}. It is however much closer to the theoretical prediction from QCD
factorization~\cite{QCDf}, thereby indicating a solution for a long-standing puzzle. 

For the decay mode $B^0\to\rho^0\rho^0$ a $3.4\sigma$ signal is
observed by Belle, which corresponds to the branching fraction of $(1.02\pm0.30\pm0.15)\times10^{-6}$.
This result is used together with results from BaBar and Belle for the other $B\to \rho\rho$ modes to obtain the value
$\alpha|_{\rho\rho}=(89.9^{+5.4}_{-5.3})^{\circ}$~\cite{Charles:2004jd}.

\begin{figure}[h]
 \centering
 \includegraphics[width=.32\textwidth]{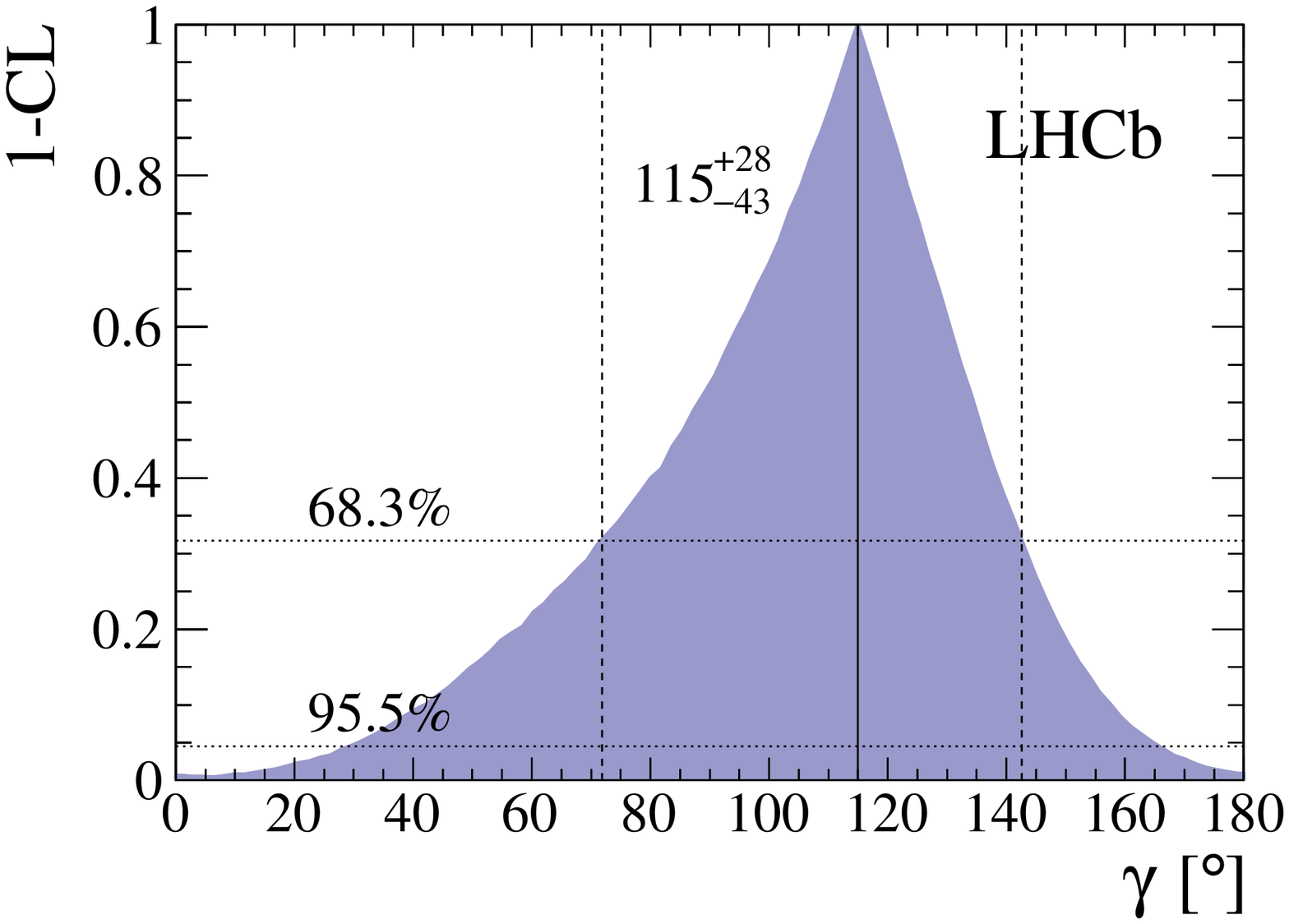}
 \includegraphics[width=.32\textwidth]{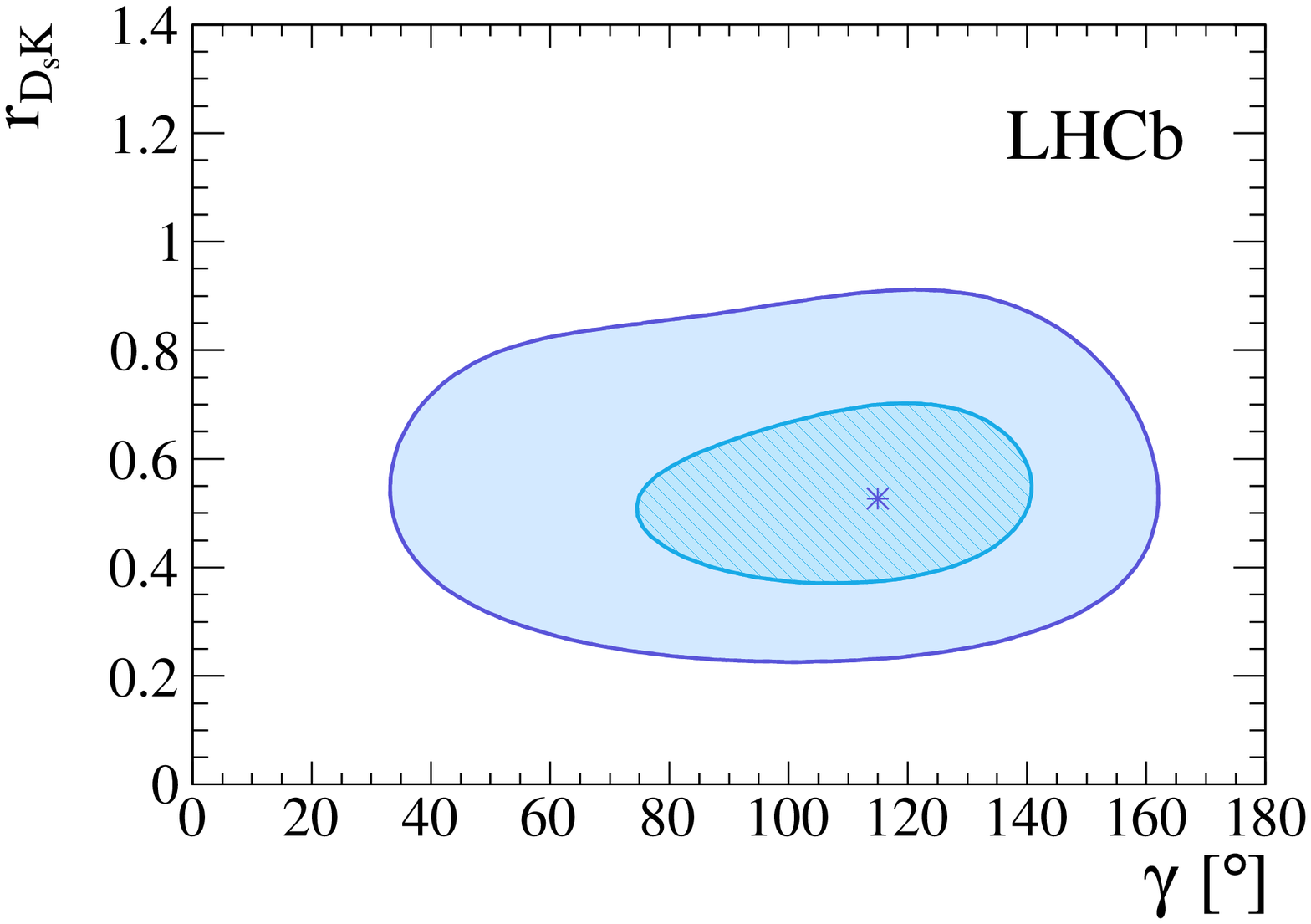}
 \includegraphics[width=.32\textwidth]{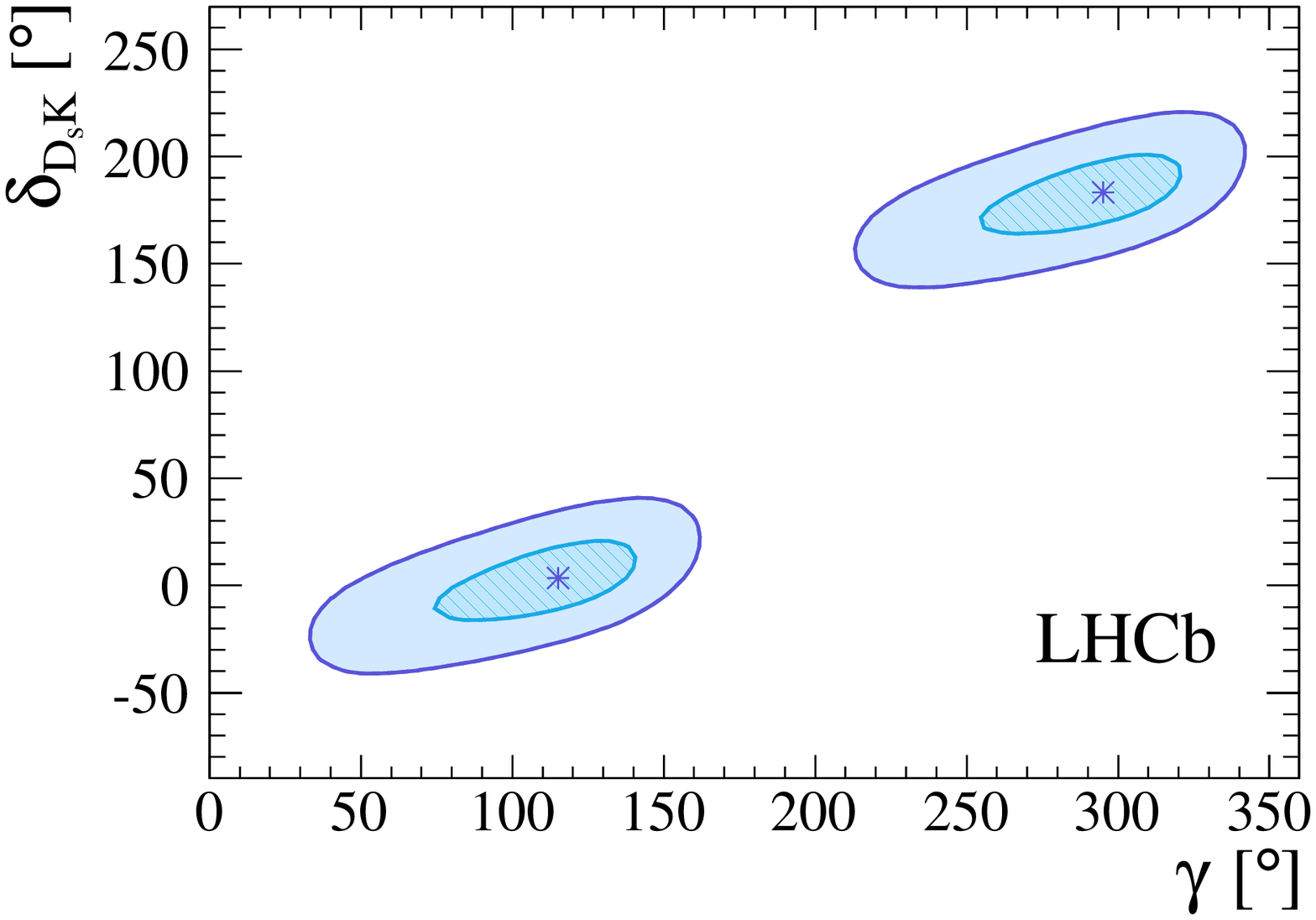}
 \caption{The SM-like solution for $\gamma$ from $B_s\to D_sK$, together with the central value and the $68.3\%$ CL interval
(left). Profile likelihood contours
of $r_{D_sK}$ vs. $\gamma$ (middle), and $\delta$ vs. $\gamma$ (right). The contours are at $1\sigma$ ($2\sigma$),
corresponding to $39\%$ CL ($86\%$ CL) in the Gaussian
approximation. The markers denote the best-fit values.
}
 \label{fig:gamma}
\end{figure}

The weak phase $\gamma$ remains the angle in the unitarity triangle with the largest uncertainties. So far it has been measured with time-independent
methods using $B^{0/+}$ decays collected at the $B$-factories and the LHCb experiment, discussed in detail in the summary of
WG~III~\cite{Libby:2014cxa}.
The focus here is on the time-dependent analysis of $B_s^0\rightarrow D_{s}^{\mp}K^{\pm}$~\cite{Fleischer:2003yb}, carried out for the first
time using $1~{\rm fb}^{-1}$ collected by the LHCb experiment~\cite{Gligorov:2014zka}. The final state is accessible from both $B_s$ and $\bar B_s$
which makes it sensitive to the combination $\phi_s+\gamma$ via the CP-violating observables $C_{{\rm CP},f}$, $A_f^{\Delta\Gamma}$,
$A_{\bar{f}}^{\Delta\Gamma}$, $S_{{\rm CP},f}$ and $S_{{\rm CP},\bar f}$, see Ref.~\cite{Gligorov:2014zka} for further details. Using the
measurement of $\phi_s$ discussed in the previous section, these observables allow for the determination of the weak phase $\gamma=
(115^{+28}_{-43})^{\circ}$, the strong phase difference between the $B_s$ and $\bar B_s$ decay $\delta= (3^{+19}_{-20})^{\circ}$ and the ratio of the absolute values of these amplitudes, $r_{D_sK}=
0.53^{+0.17}_{-0.16}$, as shown in Fig.~\ref{fig:gamma}. The phases are extracted with a two-fold ambiguity, the values quoted here correspond to the
SM-like solution. 

\section{Conclusions}

The $B$ factories, Tevatron and LHC experiments have established the validity of the Kobayashi-Maskawa mechanism of CP violation to an unexpected high
level of precision; at the moment the global picture is beautifully consistent, with very few exceptions.
This situation requires the control of uncertainties at an unprecedented level, posing new challenges for experiments and theory alike. Nevertheless,
the prospects for continued progress are excellent, as explicitly discussed
in Refs.~\cite{TalkDescotesGenon,TalkBona,GBorissov:2014,JPazzini:2014,ERodrigues:2014,Koppenburg:2014ara,TalkUrquijo} and furthermore visible in the
theory strategies presented at this workshop. Experimentally, the recently started Run~II of the LHC will dominate during the time until the next CKM
workshop. While some of the measurements are met already with corresponding theory uncertainties at an appropriate level, see \emph{e.g.} the table in
Ref.~\cite{ERodrigues:2014}, others require progress in Lattice QCD, the prospects of which have been discussed in Ref.~\cite{TalkElKhadra}, or other
non-perturbative methods, as discussed \emph{e.g.} in Sec.~\ref{sec::phids}. This combined effort visible at this workshop has the potential to
finally allow for a first glimpse of what lies beyond the SM.

\Acknowledgements
The authors would like to thank all the speakers of this working group for their contributions, all participants for stimulating discussions, and
the organizers for an enjoyable and perfectly organized workshop.


\begin{thebibliography}{99}


\bibitem{CKM}
  N.~Cabibbo,
  Phys.\ Rev.\ Lett.\  {\bf 10} (1963) 531.
  
  M.~Kobayashi and T.~Maskawa,
  Prog.\ Theor.\ Phys.\  {\bf 49} (1973) 652.

\bibitem{Charles:2004jd}
  J.~Charles {\it et al.}  [CKMfitter Group Collaboration],
  Eur.\ Phys.\ J.\ C {\bf 41} (2005) 1
  [hep-ph/0406184]. Updated results and plots available at: {\tt http://ckmfitter.in2p3.fr}.
%
\bibitem{Ciuchini:2000de}
  M.~Ciuchini, G.~D'Agostini, E.~Franco, V.~Lubicz, G.~Martinelli, F.~Parodi, P.~Roudeau and A.~Stocchi,
  JHEP {\bf 0107} (2001) 013
  [hep-ph/0012308]. Updated results and plots available at: {\tt http://www.utfit.org}.

\bibitem{Bevan:2014cya}
  A.~Bevan, M.~Bona, M.~Ciuchini, D.~Derkach, E.~Franco, V.~Lubicz, G.~Martinelli and F.~Parodi {\it et al.},
  these proceedings,
  arXiv:1411.7233 [hep-ph].


\bibitem{TalkDescotesGenon}
  S.~Descotes-Genon,
  these proceedings.

\bibitem{TalkBona}
  M.~Bona, these proceedings.
  
\bibitem{SummaryWG2}
  D.~Becirevic, M.~Della Morte and A.~Sibidanov, 
  these proceedings.



\bibitem{Lenz:2014jya}
  A.~Lenz, these proceedings,
  arXiv:1409.6963 [hep-ph].


\bibitem{Inami:1980fz}
  T.~Inami and C.~S.~Lim,
  Prog.\ Theor.\ Phys.\  {\bf 65} (1981) 297
   [Erratum-ibid.\  {\bf 65} (1981) 1772].

\bibitem{Buras:1990fn}
  A.~J.~Buras, M.~Jamin and P.~H.~Weisz,
  Nucl.\ Phys.\ B {\bf 347} (1990) 491.
  

\bibitem{Carrasco:2014nda}
  N.~Carrasco,
  arXiv:1410.0161 [hep-lat].
  
\bibitem{Lenz:2014jha}
  A.~Lenz,
  arXiv:1405.3601 [hep-ph].

\bibitem{TalkDordei}
  F.~Dordei for the LHCb collaboration,
  these proceedings,
  arXiv:1411.4247 [hep-ex].


\bibitem{LHCb:1307.2476}
R.~Aaij  {\it et al.}  [LHCb Collaboration], Phys. Rev. Lett. 111, 102003 (2013).

\bibitem{arXiv:1011.1096}
R. Fleischer and R. Knegjens, Eur. Phys. J. C71 (2011) 1532.


\bibitem{LSDA}
  V.~M.~Abazov {\it et al.}  [D0 Collaboration],
  Phys.\ Rev.\ D {\bf 82} (2010) 032001
  [arXiv:1005.2757 [hep-ex]].
  Phys.\ Rev.\ Lett.\  {\bf 105} (2010) 081801
  [arXiv:1007.0395 [hep-ex]].
  Phys.\ Rev.\ D {\bf 84} (2011) 052007
  [arXiv:1106.6308 [hep-ex]].

\bibitem{Abazov:2013uma}
  V.~M.~Abazov {\it et al.}  [D0 Collaboration],
  Phys.\ Rev.\ D {\bf 89} (2014) 1,  012002
  [arXiv:1310.0447 [hep-ex]].
  
\bibitem{TalkNierste}
  U.~Nierste,
  these proceedings.

  
\bibitem{Nebot:2014yqa}
  M.~Nebot,
  these proceedings,
  arXiv:1411.4026 [hep-ph].

\bibitem{TalkBertram}
  I.~Bertram for the D0 collaboration,
  these proceedings.


\bibitem{LHCb:1402.2554}
R.~Aaij  {\it et al.}  [LHCb Collaboration], 10.1007 JHEP 04 (2014)114, arXiv:1402.2554 [hep-ex].

  \bibitem{TGershon2010}
 T.~Gershon,
 J Phys g38 015009, 2011,
  arXiv:1007.5135 [hep-ph] 

  \bibitem{Agashe:2014kda} 
  K.~A.~Olive {\it et al.}  [Particle Data Group Collaboration],
  Chin.\ Phys.\ C {\bf 38}, 090001 (2014).

\bibitem{LHCb:arXiv:1405.7223}
R.~Aaij  {\it et al.}  [LHCb Collaboration], Phys. Rev. Lett. 113, 032001 (2014), arXiv:1405.7223[hep-ex].

\bibitem{LHCb:arXiv:1405.1543}
R.~Aaij  {\it et al.}  [LHCb Collaboration], Phys. Lett. B 736, 154-162 (2014), arXiv:1405.1543 [hep-ex].

\bibitem{Fleischer:2011cw}
  R.~Fleischer and R.~Knegjens,
  Eur.\ Phys.\ J.\ C {\bf 71} (2011) 1789
  [arXiv:1109.5115 [hep-ph]].

\bibitem{Jung:2014jfa}
  M.~Jung and S.~Schacht,
  Phys.\ Rev.\ D {\bf 91} (2015) 3,  034027
  [arXiv:1410.8396 [hep-ph]].

\bibitem{LHCb:arXiv:1407.5873}
R.~Aaij  {\it et al.}  [LHCb Collaboration], Phys. Rev. Lett. 113, 172001 (2014),  arXiv:1407.5873 [hep-ex].
\bibitem{D0:arXiv:1410.1568}
V.~ M. Abazov {\it et al.}  [D0 Collaboration], Phys. Rev. Lett. 114, 062001, arXiv:1410.1568 [hep-ex].

\bibitem{Lenz:2006hd}
  A.~Lenz and U.~Nierste,
  JHEP {\bf 0706} (2007) 072
  [hep-ph/0612167].

\bibitem{NPBmixing}
  M.~Bona {\it et al.}  [UTfit Collaboration],
  JHEP {\bf 0803} (2008) 049
  [arXiv:0707.0636 [hep-ph]].
  A.~Lenz, U.~Nierste, J.~Charles, S.~Descotes-Genon, A.~Jantsch, C.~Kaufhold, H.~Lacker and S.~Monteil {\it et al.},
  Phys.\ Rev.\ D {\bf 83} (2011) 036004
  [arXiv:1008.1593 [hep-ph]].
  A.~Lenz, U.~Nierste, J.~Charles, S.~Descotes-Genon, H.~Lacker, S.~Monteil, V.~Niess and S.~T'Jampens,
  Phys.\ Rev.\ D {\bf 86} (2012) 033008
  [arXiv:1203.0238 [hep-ph]].

\bibitem{Bobeth:2011st}
  C.~Bobeth and U.~Haisch,
  Acta Phys.\ Polon.\ B {\bf 44} (2013) 127
  [arXiv:1109.1826 [hep-ph]].
  
\bibitem{Bobeth:2014rda}
  C.~Bobeth, U.~Haisch, A.~Lenz, B.~Pecjak and G.~Tetlalmatzi-Xolocotzi,
  JHEP {\bf 1406} (2014) 040
  [arXiv:1404.2531 [hep-ph]].

   
\bibitem{LGrillo:2014}
L.~Grillo, 
these proceedings,
 arXiv:1411.6198 [hep-ex]

\bibitem{LHCb:arXiv:1409.8586}
R.~Aaij  {\it et al.}  [LHCb Collaboration], Phys. Rev. Lett. 114, 041601 (2015), arXiv:1409.8586 [hep-ex].

\bibitem{TalkCheng}
  C.-H.~Cheng, these proceedings.

\bibitem{Lees:2014kep}
  J.~P.~Lees {\it et al.} [BaBar Collaboration],
  Phys.\ Rev.\ Lett.\  {\bf 114} (2015) 8,  081801
  [arXiv:1411.1842 [hep-ex]].

\bibitem{Borissov:2013wwa}
  G.~Borissov and B.~Hoeneisen,
  Phys.\ Rev.\ D {\bf 87} (2013) 7,  074020
  [arXiv:1303.0175 [hep-ex]].
  
\bibitem{Botella:2014qya}
  F.~J.~Botella, G.~C.~Branco, M.~Nebot and A.~Sanchez,
  arXiv:1402.1181 [hep-ph].
  
  


  
\bibitem{Bigi:1981qs}
  I.~I.~Y.~Bigi and A.~I.~Sanda,
  Nucl.\ Phys.\ B {\bf 193} (1981) 85.

\bibitem{PenEstimates}  
  H.~Boos, T.~Mannel and J.~Reuter,
  Phys.\ Rev.\ D {\bf 70} (2004) 036006
  [hep-ph/0403085].
  %
  H.~-n.~Li and S.~Mishima,
  JHEP {\bf 0703} (2007) 009
  [hep-ph/0610120].
  %
  M.~Gronau and J.~L.~Rosner,
  Phys.\ Lett.\ B {\bf 672} (2009) 349
  [arXiv:0812.4796 [hep-ph]].
  X.~Liu, W.~Wang and Y.~Xie,
  Phys.\ Rev.\ D {\bf 89} (2014) 094010
  [arXiv:1309.0313 [hep-ph]].

\bibitem{Aaij:2014zsa}
  R.~Aaij {\it et al.} [LHCb Collaboration],
  Phys.\ Rev.\ Lett.\  {\bf 114} (2015) 4,  041801
  [arXiv:1411.3104 [hep-ex]].
\bibitem{Aad:2014cqa}
  G.~Aad {\it et al.}  [ATLAS Collaboration],
  Phys.\ Rev.\ D {\bf 90} (2014) 5,  052007
  [arXiv:1407.1796 [hep-ex]].
\bibitem{Khachatryan:2015nza}
  V.~Khachatryan {\it et al.} [CMS Collaboration],
  arXiv:1507.07527 [hep-ex].


\bibitem{Fleischer:1999nz}
  R.~Fleischer,
  Eur.\ Phys.\ J.\ C {\bf 10} (1999) 299
  [hep-ph/9903455].
  
\bibitem{PenUspin}
  R.~Fleischer,
  Eur.\ Phys.\ J.\ C {\bf 10} (1999) 299
  [hep-ph/9903455].
 %
  M.~Ciuchini, M.~Pierini and L.~Silvestrini,
  Phys.\ Rev.\ Lett.\  {\bf 95} (2005) 221804
  [hep-ph/0507290];
 %
  arXiv:1102.0392 [hep-ph].
%
\bibitem{Faller:2008zc}
  S.~Faller, M.~Jung, R.~Fleischer and T.~Mannel,
  Phys.\ Rev.\ D {\bf 79} (2009) 014030
  [arXiv:0809.0842 [hep-ph]].
%
\bibitem{Faller:2008gt}
  S.~Faller, R.~Fleischer and T.~Mannel,
  Phys.\ Rev.\ D {\bf 79} (2009) 014005
  [arXiv:0810.4248 [hep-ph]].

\bibitem{Feldmann:2008fb}
  T.~Feldmann, M.~Jung and T.~Mannel,
  JHEP {\bf 0808} (2008) 066
  [arXiv:0803.3729 [hep-ph]].

\bibitem{Jung:2012mp}
  M.~Jung,
  Phys.\ Rev.\ D {\bf 86} (2012) 053008
  [arXiv:1206.2050 [hep-ph]].

\bibitem{Gronau:2008hb}
  M.~Gronau and J.~L.~Rosner,
  Phys.\ Lett.\ B {\bf 669} (2008) 321
  [arXiv:0808.3761 [hep-ph]].
  
  
\bibitem{TalkFrings}
  P.~Frings, 
  these proceedings.  

\bibitem{Frings:2015eva}
  P.~Frings, U.~Nierste and M.~Wiebusch,
  arXiv:1503.00859 [hep-ph].

\bibitem{BtoJPsif0prop}
  S.~Stone and L.~Zhang,
  Phys.\ Rev.\ D {\bf 79} (2009) 074024
  [arXiv:0812.2832 [hep-ph]].
%
  arXiv:0909.5442 [hep-ex].

\bibitem{ReviewsScalars}
  E.~Klempt and A.~Zaitsev,
  Phys.\ Rept.\  {\bf 454} (2007) 1
  [arXiv:0708.4016 [hep-ph]].
Amsler {\it et al.} in
  K.~A.~Olive {\it et al.}  [Particle Data Group Collaboration],
  Chin.\ Phys.\ C {\bf 38} (2014) 090001.
  
\bibitem{Fleischer:2011au}
  R.~Fleischer, R.~Knegjens and G.~Ricciardi,
  Eur.\ Phys.\ J.\ C {\bf 71} (2011) 1832
  [arXiv:1109.1112 [hep-ph]].
  
\bibitem{Aaij:2014siy}
  R.~Aaij {\it et al.}  [LHCb Collaboration],
  Phys.\ Rev.\ D {\bf 90} (2014) 1,  012003
  [arXiv:1404.5673 [hep-ex]].

\bibitem{Knegjens:2014fqa}
  R.~Knegjens,
  these proceedings,
  arXiv:1411.3980 [hep-ph].

   
   \bibitem{GBorissov:2014}
   G.~Borissov for the ATLAS Collaboration, 
   these proceedings.
     
   \bibitem{JPazzini:2014} 
   J.~Pazzini for the CMS Collaboration,
   these proceedings, 
   arXiv:1411.3560 [hep-ex].
   
  \bibitem{WKanso:2014}
  W.~Kanso for the LHCb Collaboration, 
  these proceedings, 
  arXiv:1411.5918 [hep-ex].

\bibitem{Aaij:2013oba}
  R.~Aaij {\it et al.} [LHCb Collaboration],
  Phys.\ Rev.\ D {\bf 87} (2013) 11,  112010
  [arXiv:1304.2600 [hep-ex]].

\bibitem{Aaij:2014dka}
  R.~Aaij {\it et al.} [LHCb Collaboration],
  Phys.\ Lett.\ B {\bf 736} (2014) 186
  [arXiv:1405.4140 [hep-ex]].

\bibitem{Aaij:2013eia}
  R.~Aaij {\it et al.} [LHCb Collaboration],
  Nucl.\ Phys.\ B {\bf 873} (2013) 275
  [arXiv:1304.4500 [hep-ex]].

\bibitem{Aaij:2012nh}
  R.~Aaij {\it et al.} [LHCb Collaboration],
  Phys.\ Rev.\ D {\bf 86} (2012) 071102
  [arXiv:1208.0738 [hep-ex]].

\bibitem{Aaij:2015mea}
  R.~Aaij {\it et al.} [LHCb Collaboration],
  arXiv:1509.00400 [hep-ex].


\bibitem{TalkSchacht}
 St.~Schacht, these proceedings.


\bibitem{CFitzpatrick:2014}
  C.~Fitzpatrick for the LHCb Collaboration, 
  these proceedings. 

\bibitem{Aaij:2014ywt}
  R.~Aaij {\it et al.} [LHCb Collaboration],
  Phys.\ Rev.\ Lett.\  {\bf 113} (2014) 21,  211801
  [arXiv:1409.4619 [hep-ex]].

\bibitem{2dstbabar}
J.P.~Lees  {\it et al.}  (The BaBar Collaboration),
Phys. Rev. D {\bf 86}, 112006 (2012).

\bibitem{TMiyashita}
   T.~Miyashita, these proceedings.

\bibitem{Beneke:2005pu}
  M.~Beneke,
  Phys.\ Lett.\ B {\bf 620} (2005) 143
  [hep-ph/0505075].

\bibitem{TalkRitter}
  M.~Ritter,
  these proceedings.

\bibitem{omegaKbelle}
V.~Chobanova {\it et al.} (The Belle Collaboration),
Phys.Rev.D 90, 012002 (2014).

\bibitem{etaprimeKbelle}
L.~Santelj {\it et al.} (The Belle Collaboration),
JHEP 1410, 165 (2014).

\bibitem{gronau}
M. Gronau and D. London, 
Phys. Rev. Lett. 65, 3381 (1990).



  







\bibitem{Vanhoefer:2014mfa}
  P.~Vanhoefer for the Belle Collaboration, these proceedings,
  arXiv:1410.5700 [hep-ex].

\bibitem{Libby:2014cxa}
  J.~Libby, these proceedings,
  arXiv:1412.4269 [hep-ex].

\bibitem{Abe:2004mp}
  K.~Abe {\it et al.} [Belle Collaboration],
  Phys.\ Rev.\ Lett.\  {\bf 94} (2005) 181803
  [hep-ex/0408101].

\bibitem{Lees:2012mma}
  J.~P.~Lees {\it et al.} [BaBar Collaboration],
  Phys.\ Rev.\ D {\bf 87} (2013) 5,  052009
  [arXiv:1206.3525 [hep-ex]].


\bibitem{QCDf}
  M.~Beneke, G.~Buchalla, M.~Neubert and C.~T.~Sachrajda,
  Nucl.\ Phys.\ B {\bf 606} (2001) 245
  [hep-ph/0104110].

  M.~Beneke and M.~Neubert,
  Nucl.\ Phys.\ B {\bf 675} (2003) 333
  [hep-ph/0308039].


\bibitem{Fleischer:2003yb}
  R.~Fleischer,
  Nucl.\ Phys.\ B {\bf 671} (2003) 459
  [hep-ph/0304027].

\bibitem{Gligorov:2014zka}
  V.~Gligorov for the LHCb Collaboration, these proceedings,
  arXiv:1411.4865 [hep-ex].
  
\bibitem{ERodrigues:2014}
E.~Rodrigues,  
these proceedings. 

\bibitem{Koppenburg:2014ara}
  P.~Koppenburg, these proceedings,
  arXiv:1411.5119 [hep-ex].

\bibitem{TalkUrquijo}
  P.~Urquijo, these proceedings.
  
\bibitem{TalkElKhadra}
  A.~El-Khadra, these proceedings.
\end{thebibliography}
\end{document}